%Paper: hep-th/9406201
%From: Michael McGuigan <mcguigan@phys.ufl.edu>
%Date: Wed, 29 Jun 1994 18:32:44 -0400

\tolerance=5000
\input phyzzx
\nopubblock
\titlepage
\line{\hfill UFIFT-HEP-94-07}
\line{\hfill }
\line{\hfill }
\title{Finite Black Hole Entropy and String Theory}
\author{Michael McGuigan\foot{Research supported in part by the U.S.
Department of
     Energy under contract no. DE-FG05-86ER-40272. } }
\vskip.2cm
\centerline{\it Institute for Fundamental Theory}
\centerline{\it University of Florida}
\centerline{\it Gainesville, Fl. 32611 }
\centerline{\it E-mail : mcguigan@phys.ufl.edu}

\abstract{An accelerating observer sees a thermal bath of radiation at the
Hawking
temperature which is proportional to the acceleration. Also, in string
theory there is a Hagedorn temperature beyond
which one cannot go  without an infinite amount of energy. Several authors
have shown that in the context of Hawking radiation a limiting temperature
for string theory leads to a limiting acceleration, which for a black hole
implies a minimum distance from the horizon for an observer to remain
stationary.
We argue that this effectively introduces a cutoff in Rindler space
or the Schwarzschild
geometry inside of which accelerations would exceed this maximum value.
Furthermore, this natural
cutoff in turn allows one to define a finite entropy for Rindler space or
a black hole as all divergences were occurring on the horizon. In all cases
if a particular relationship exists between Newton's constant and the string
tension then the entropy of the string modes agrees with the
Bekenstein-Hawking formula.
}
\endpage

\REF\thoo{G.t'Hooft, Nucl.Phys.B256,727 (1985)}
\REF\bomb{L.Bombelli, R.K.Koul, J.Lee and R.Sourkin, Phys.Rev.D34,373 (1986)}
\REF\sred{M.Srednicki, Phys.Rev.Lett.71, 666 (1993)}
\REF\call{C.Callan and F.Wilczek, Inst.Adv.Study Preprint IASSNS-HEP-93-87,
hep-th@xxx.lanl.gov-9401072}
\REF\kaba{D.Kabat and M.Strassler, Rutgers Preprint RU-94-10, Jan.1994,
hep-th@xxx.lanl.gov-9401125}
\REF\suss{L.Susskind and J.Uglum, Stanford Preprint SU-ITP-94-1, Jan.1994,
hep-th@xxx.lanl.gov-9401070}
\REF\sussk{\L.Susskind, Rutgers Preprint RU-93-44, Oct.1993,
hep-th@xxx.lanl.gov-
9309145}
\REF\russ{J.Russo and L.Susskind, Texas Preprint UTTG-9-94, May 1994,
hep-th@xxx.lanl.gov-9405117}
\REF\dowk{J.Dowker, Class.Quant.Grav.11,L55-L60, (1994)}
\REF\hage{R.Hagedorn, Nuovo Cimento Suppl.3, 147 (1965); 6,311 (1968);
Nuovo Cimento 52A, 1336 (1967); 56A, 1027 (1968)}
\REF\huan{K.Huang and S.Weinberg, Phys.Rev.Lett.25,895 (1970)}
\REF\hawk{S.Hawking, Comm.Math.Phys.43,199 (1975)}
\REF\saka{N.Sakai, "Hawking Radiation in String Theories"
in honor of 60th Birthday of Yoshio Yamaguchi,
Particles and Nuclei:286 (1986)}
\REF\pare{R.Parentani and R.Potting, Phys.Rev.Lett.63,945 (1989)}
\REF\bowi{M.J.Bowick and S.B.Giddings, Nucl.Phys.B325, 631 (1989)}
\REF\bowic{M.J.Bowick and L.C.R. Wijewardhana, Phys.Rev.Lett.54, 2485 (1985)}
\REF\alva{E.Alvarez, Phys.Rev.D31, 418 (1985); Nucl.Phys.B269, 596 (1986)}
\REF\anto{I.Antoniadis, J.Ellis and D.V.Nanopoulos, Phys.Lett.B199, 402 (1987)}
\REF\birr{N.D.Birrell and P.C.W.Davies,"Quantum Fields in Curved Space",
Cambridge Univ. Press (1982)}
\REF\deve{H.J.Devega and N.Sanchez, Nucl.Phys.B299, 818 (1988);
Nucl.Phys.B349, 815 (1991)}
\REF\feol{A.Feoli, Nucl.Phys.B396, 261 (1993)}
\REF\cand{P.Candelas and D.Deutsch,
Proc.Roy.Soc.Lond.A354,79 (1977); A362,251 (1978)}
\REF\taka{S.Takagi, Prog.Theor.Phys.72, 505 (1984);
Prog.Theor.Phys.74, 142,501,1219 (1985)}
\REF\isra{W.Israel, Phys.Lett.57A, 107 (1976)}
\REF\gros{D.J.Gross, J.Harvey, E.Martinec and R.Rohm, Nucl.Phys.B267,75 (1986)}
\REF\schw{J.Schwarz, Phys.Rep.89, 223 (1982); M.Green, J.Schwarz and
E.Witten,"Supersting Theory", Cambridge Univ. Press (1987)}
\REF\frau{S.Frautchi, Phys.Rev.D3,2821 (1971); S.Carlitz,
Phys.Rev.D5,3231 (1972)}
\REF\turo{N.Turok, PhysicaA158,516 (1989)}
\REF\bowick{M.Bowick, L.Smolin and L.C.R.Wijewardhana, Phys.Rev.Lett.56,424
(1986)}
\REF\cohe{A.Cohen, G.Moore, P.Nelson and J.Polchinski,
Nucl.Phys.B267,143 (1986)}

\noindent {\bf I. Introduction}

Usually entropy in physics can be described both in a thermodynamic sense
and in terms of a counting of states. However although black hole entropy
has long been formulated in a thermodynamic sense involving the Hawking
temperature, only recently has it been possible to approach black hole
entropy as a counting of states. In \NPrefmark{\thoo} t'Hooft
calculated the number
of particle states surrounding a black hole in a "brick wall model"
where particles are not allowed to be closer than a certain cutoff
distance to the horizon. He found a contribution to the entropy
proportional to the area of the horizon but divergent as the cutoff
distance was taken to zero. A different approach was taken by
Bombelli et al \NPrefmark{\bomb} and Srednicki \NPrefmark{\sred}
who traced over particle states inside the sphere
of the horizon and also found a divergent entropy proportional to
the area. Callan and Wilczek \NPrefmark{\call} and
Kabat and Strassler \NPrefmark{\kaba} showed that the
brick wall model of t'Hooft
and the geometric model of \NPrefmark{\bomb, \sred}
were in fact equivalent. The divergence in the entropy arises
because of an infinite number of states which occur on the horizon
itself and occurs whenever the brick wall cutoff is removed. Susskind and
Uglum \NPrefmark{\suss}
calculated the density of states of the Schwarzschild geometry
in the limit
of infinite mass, which was essentially equivalent to Rindler space,
the spacetime seen by an accelerating observer. The canonical particle
entropy was again divergent but they identified a
contribution in string theory consisting of open
strings with the ends attached to the horizon that, owing to the different
ultraviolet properties of string theory, could in principle yield
a finite black hole entropy \NPrefmark{\sussk, \russ, \dowk}.

In this paper we pursue an alternate route to a finite entropy in superstring
theory. We show that the brick wall cutoff used by t'Hooft to obtain a finite
entropy has a natural interpretation in terms of a string theory's
maximum acceleration. This is fundamentally a string theory phenomena related
to
the existence of a Hagedorn temperature \NPrefmark{\hage} \NPrefmark{\huan},
the limiting temperature in string
theory. Essentially an accelerating observer sees thermal radiation
at the Hawking
temperature $T = {a \over 2\pi }$ with $a$ the acceleration \NPrefmark{\hawk}.
In string
theory however there is a limiting temperature, the Hagedorn temperature, this
suggests that
for Hawking radiation there is a limiting Hawking temperature and maximum
acceleration. Sakai \NPrefmark{\saka} has studied this phenomena by calculating
the thermal response function and vacuum stress in Rindler space and
finds a limiting Hawking temperature related to the Hagedorn temperature by
$T_{Hawking-Max} = T_{Hagedorn}/\pi$. Parentani and Potting \NPrefmark{\pare}
find the same relation in
terms of a thermal Greens function approach (see in addition Bowick and
Giddings
\NPrefmark{\bowi}).
This limiting acceleration also applies to stationary observers outside a
black hole  and yields minimum distance from the horizon for
an observer to remain stationary. In this paper we find that
the structure of Rindler
space or the Schwarzschild geometry must be suitably altered close to
the horizon, where accelerations would exceed the maximum value and infinite
vacuum stress would be present.  We then argue that this effectively introduces
a cutoff in Rindler space or the Schwarzschild geometry and yields
a finite entropy in string theory.

\noindent{\bf II. Maximum Acceleration in String Theory}

The essence of the Hawking effect is that an observer at
constant acceleration $a$ will feel
the existence of a heat bath at temperature $T= {a\over 2\pi}$. It is also well
known that in string theory there exist a maximum temperature $T_{Hagedorn}$
above
which the string partition function diverges. Essentially this is because
the number of string states of given mass is
$\rho (m) = m_0^{n_s -1}m^{-n_s} \exp(bm)$
and the partition function involves multiplying by $\exp{(-\beta m)}$
and summing over $m$, where $\beta = 1/T$. Thus the critical temperature
is given by $T_{Hagedorn} = 1/b$. One might expect that because the Hawking
radiation is thermal, string theories possess a limiting Hawking temperature
corresponding to $1/b$. Indeed Sakai \NPrefmark{\saka} and also Parentani
and Potting \NPrefmark{\pare}
showed that there is maximum Hawking temperature, however the limiting
value turns out to be $T_{Hawking-Max} = 1/b\pi$ so that the limiting
acceleration is simply $a_{max} = 2/b$. For the mass degeneracy
$\rho(m) = m_0^{n_s-1}m^{-n_s} \exp{(bm)}$ the quantities $n_s$ and
$b$ are determined by the spectrum of various string theories
\NPrefmark{\bowic}
\NPrefmark{\alva} \NPrefmark{\anto}. One has for $n$ noncompact dimensions
$n_s= {n+1\over 2}$ for open strings, $n_s= n$ for closed strings,
$b$ is given by $4\pi \sqrt{\alpha'}$ for bosonic strings,
$2\sqrt{2}\pi\sqrt {\alpha'}$ for superstrings and
$(2+\sqrt{2})\pi\sqrt{\alpha'}$
for heterotic superstrings
while $m_0$ is of the order $1/\sqrt{\alpha'}$

Before discussing the derivation of this maximum acceleration let us
quickly review the structure of Rindler space \NPrefmark{\birr}, the
spacetime as seen from an accelerated observer,
in order to set notation useful later.
Solving the differential equation
$$
{d \over dt} ( {1 \over \sqrt{1-v^2}} {dx\over dt} ) = a
\eqno(2.1)$$
with $v = {dx\over dt}$ and $x_{\perp}=0$, we obtain the spacetime trajectory
of a particle with proper acceleration $a$, initial position and velocity
$x_i$, $v_i$ at time $t_i$ given by:
$$
x = x_i + {1\over a} \Biggl(
\sqrt{1+ (a(t-t_i)+ {v_i\over \sqrt{1-v_i^2} })^2 }
- \sqrt{1+ {v_i^2 \over 1-v_i^2}} \Biggr)
.\eqno(2.2)$$
Now if the relations
$t_i = {1\over a} {v_i \over \sqrt{1-v_i^2} }$
and
$x_i = {1\over a}\sqrt{1+ {v_i^2 \over 1-v_i^2}}$
hold for the initial conditions, the
spacetime trajectory simplifies dramatically to
$$x= \sqrt{ t^2 + {1\over a^2} }.\eqno(2.3)$$
This is the trajectory of a Rindler observer. Note that the distance
of closest approach to the origin $x_{min} = 1/a$ is smaller for
larger accelerations, whereas one might expect large accelerations
to cause a turn around further out. This happens because the
initial conditions of a Rindler
trajectory are such that highly accelerated observers start closer
to the origin. The Rindler coordinates
are defined by $x= s\cosh{\tau}$, $t=s\sinh{\tau}$, and cover the
righthand wedge of Minkowski space $x > |t|$. Then from (2.3) we have
the relation
$$s= 1/a,\eqno(2.4)$$
so that large accelerations correspond to small $s$ close to the horizon
$x = |t|$. In terms of Rindler coordinates the flat spacetime metric
takes the form\hfill\break
$d\ell^2 = -s^2d\tau^2 + ds^2 + dx_{\perp}dx_{\perp}$. As described by
Susskind and Uglum the Schwarzschild metrc
$d\ell^2=- (1- {2GM\over r})dt^2 + (1- {2GM\over r})^{-1}dr^2 + d\Omega^2$
of a very large black hole can also be described by the Rindler
coordinates with the transformation $\tau= {t\over 4GM}$,
$s= \sqrt{8GM(r-2GM)}$ and the area of the
horizon taken to be $A = (2GM)^2$.

The phenomena of maximum acceleration in string theory has been discussed
in several contexts. Sakai \NPrefmark{\saka} studied the
detector response functional
for string theories in an accelerating frame and found that it diverged
for $a > 2/b= a_{max}$. Parentani and Potting \NPrefmark{\pare} studied the
Feynman propagator
in Rindler space and found the same value for the limiting acceleration.
Another approach regarding acceleration in string theory was taken
in \NPrefmark{\deve, \feol} where a mode instability for
classical solutions of an extended object
in Rindler space lead to a critical acceleration
$a_c = \bigl({3\over \pi (n-2)}\bigl)^{1\over 3} {1\over \sqrt{\alpha'}}$.
Because the critical acceleration $a_c$ is somewhat larger than the
maximum acceleration $a_{max}= 2/b$ of Sakai, we mainly work with $a_{max}$,
as these effects should occur first, however similar conclusions can also
be reached regarding $a_c$.
A physically intuitive derivation of the maximum acceleration was also
given by Sakai who studied the difference in the vacuum energy between
Rindler space and Minkowski space, similar to the Casimir effect. One begins
with the stress tensor for a massive scalar field
$$T_{\mu\nu}= \partial_\mu \phi \partial_\nu \phi
   - {1\over 2} g_{\mu \nu}
( g^{\rho \sigma}\partial_\rho \phi \partial_\sigma \phi
+ m^2 \phi^2 ).\eqno(2.5)$$
The vacuuum stress \NPrefmark{\cand} \NPrefmark{\taka} is then defined by
$\bar{T}_{\mu\nu}= \langle 0_R| T_{\mu\nu}|0_R\rangle
- \langle 0_M| T_{\mu\nu}|0_M\rangle $
where $| 0_R\rangle$ and $|0_M\rangle$ represent the Rindler and Minkowski
vacuum respectively.

The vacuum stress can
be computed by solving the eigenvalue problem in Rindler and
Minkowski space and using these solutions to form the two point function
and then the stress tensor from
$$\bar{T}_{\mu\nu}=\lim_{x\rightarrow x'} \Biggl(\partial_\mu  \partial_\nu'
   - {1\over 2} g_{\mu \nu}
( g^{\rho \sigma}\partial_\rho  \partial_\sigma'
+ m^2 ) \Biggr)\bar{G}(x,x')\eqno(2.6)$$
where $\bar{G}(x,x') =
\bar{G}(x,x')= \langle 0_R| \phi(x)\phi(x')|0_R\rangle
- \langle 0_M| \phi(x)\phi(x')|0_M\rangle $.
The quantum field $\phi$ is written
$$\phi = \int dw \int {d^{n-2}k \over (2\pi)^{n-2\over 2} } a_k u_k + h.c.
.\eqno(2.7)$$
Here $u_k$ are eigenfunctions in Rindler space and $k$ is the conjugate
momentum to $x_{\perp}$.
The $a_k$ annihilate the Rindler vacuum $a_k|0_R\rangle =0$ and are related
to Minkowski space creation and annihilation operators through
$a_k = \sqrt{1+n_w} d_k^{\dagger} + \sqrt{n_w} d_k$ with $d_k|0_M\rangle =0$
and $n_w = (e^{2\pi w} -1)^{-1}$ \NPrefmark{\isra}.

The Greens functions are then given by
$\langle 0_R|\phi(x)\phi(x')|0_R\rangle = \int_0^{\infty} dw
\int {d^{n-2}k \over ({2\pi})^{n-2}} u_k(x) u_k^{*}(x')$ and
$\langle 0_M|\phi(x)\phi(x')|0_M\rangle = \int_0^{\infty} dw
\int {d^{n-2}k \over ({2\pi})^{n-2}}( n_w u_k^{*}(x) u_k(x')
+ (1+n_w) u_k(x) u_k^{*}(x') )$ so that
$$\bar{G}(x,x') = -\int_0^{\infty} dw
\int {d^{n-2}k \over ({2\pi})^{n-2}} n_w u_k^{*}(x) u_k(x') + c.c..\eqno(2.8)$$
The eigenfunctions in Rindler space are solutions to:
$$(-{1\over s^2} {\partial^2 \over \partial\tau^2}
 +{\partial^2 \over \partial s^2} + {1\over s}{\partial \over \partial s}
 +{\partial \over \partial x_{\perp}}{\partial \over \partial x_{\perp}} - m^2)
u_k = 0
\eqno(2.9)$$
and one obtains:
$$u_k = {1\over \pi} (\sinh{\pi w})^{1\over 2} K_{iw}(s\sqrt{k^2 + m^2})
\exp{i(kx_{\perp}-w\tau)}\eqno(2.10)$$
with $K_{iw}(z)$ the modified Bessel function.
Now one uses this solution to compute the Greens function $\bar{G}(x,x)$
obtaining the stress tensor from (2.6). For large mass, such as the massive
states in a string theory, the last term in (2.6) is dominant and we
find for the stress energy:
$$ \bar{T}_0^0 \approx m^2\int_0^{\infty} dw (e^{2\pi w} - 1)^{-1}
{1 \over \pi^2} \sinh{\pi w}
\int {d^{n-2}k \over ({2\pi})^{n-2}} |K_{iw}(s\sqrt{k^2+m^2})|^2.\eqno(2.11)$$
The fermionic contribution to the vacuum stress has a similar form
except for a Fermi-Dirac factor $(e^{2\pi w}+1)^{-1}$\NPrefmark{\cand}.
Using the asymptotic expansion for the Bessel function
$K_{iw}(z) \approx \sqrt{{\pi \over 2z}} e^{-z}$
and integrating over $w$ and $k$ one obtains for the vacuum stress energy
at large mass
$$\bar{T}_0^0(s,m) \sim (m/s)^{{n\over 2}}e^{-2ms}\eqno(2.12)$$
which is in agreement with
Sakai \NPrefmark{\saka} and Takagi \NPrefmark{\taka}.

Now one forms the vacuum stress
for a string theory $\bar{T}_0^{0S}$ by multiplying (2.12)
by the number of string states
at a given mass $\rho(m)= m_0^{n_s -1}m^{-n_s}e^{bm}$ and integrating over the
mass to obtain
$$\bar{T}_0^{0S} = \int_{m_0}^{\infty} dm \rho(m) \bar{T}_0^0(s,m)
\sim \int_{m_0}^{\infty}
dm m_0^{n_s-1}m^{-n_s} e^{bm}(m/s)^{{n\over 2}} e^{-2ms} .\eqno(2.13)$$
This clearly diverges $s < b/2 = s_{min}$. Now since $s = 1/a$, the string
stress energy diverges
for accelerations $a > 2/b = a_{max}$ and and we have the limiting acceleration
of Sakai \NPrefmark{\saka} and Parentani and Potting \NPrefmark{\pare}.
The physical interpretation
of this result can be inferred from the work of Candelas and Deutsch
\NPrefmark{\cand} who showed that $\bar{T}_{00}$ represents the absence
of Hawking radiation from the vacuum and the presence of
$T_{00}^{thermal}= \bar{T}_0^0$ thermal energy density in Rindler space.
Therefore the divergence in $\bar{T}_{00}^S$ for $s< s_{min}$
represents the absence from
the vacuum of an infinite amount of Hawking radiation and there is a infinite
wall of thermal stress energy $T_{00}^{thermal} = \bar{T}_0^{0S}$
a finite
distance from the horizon. Since the energy of the thermal bath of radiation
is taken from the external source accelerating the observer \NPrefmark{\birr},
it follows that
a string cannot accelerate into that region of Rindler space. The energy
required to accelerate further cannot be produced and the string can only
continue at a uniform velocity into that region (see Figure 1.). Also by the
equivalence principle, a string cannot remain stationary at a distance $s<
s_{min}$ from the horizon of a Schwarzschild geometry. The energy
required to support the inifinite
thermal stress present there cannot be produced and the string simply
slips into the black hole.
\endpage
\noindent{\bf III. Maximum Acceleration and Finite Entropy}

The presence of an infinite wall of stress energy in Rindler space
introduces a forbidden region and cutoff $s> s_{min}$ which has important
implications
for finite entropy. Consider the single particle density of states of
a scalar particle of mass $m$. Susskind and Uglum \NPrefmark{\suss}
took the eigenvalue
equation (2.9) and solving for the radial momentum
$p_s = ({ w^2 \over s^2} - k^2 -m^2)^{1\over 2}$ with turnaround points
$s_{max} = {w \over \sqrt{k^2+m^2}}$ and $s_{min}$ obtained
$n\pi = \int_{s_{min}}^{s_{max}} p_s ds =
\int_{s_{min}}^{{w \over \sqrt{k^2+m^2}}}
({ w^2 \over s^2} - k^2 -m^2)^{1\over 2}ds$ for quantum number $n$.
The single particle density of states represents the Jacobian
$g(w,k,m) = {dn\over dw}$ between the
discrete index $n$ and the quantity $w$ and Susskind and Uglum find
$$
g(w,k,m) = {1\over 2\pi}
\log{ \Bigl({ {w\over s_{min}} + p_s(s_{min})\over
{w\over s_{min}} - p_s(s_{min})} \Bigr)}\eqno(3.1)$$
where $p_s(s_{min}) = ({ w^2 \over s_{min}^2} - k^2 -m^2)^{1\over 2}$.
Integrating over the transverse momenta for $n=4$ they obtain
$$g(w,m) = \int A{d^2 k \over (2\pi)^2 } g(w,k) =
{A\over (2\pi)^2}\Biggl( {w\over s_{min}}\sqrt{ {w^2 \over s_{min}^2}-m^2}
+ {m^2\over 2}
\log{\Bigl( {{w\over s_{min}}- \sqrt{ {w^2 \over s_{min}^2}-m^2}\over
{w\over s_{min}}- \sqrt{ {w^2 \over s_{min}^2}-m^2} }\Bigr)}\Biggr).
\eqno(3.2)$$
The second contribution vanishes for a massless particle but is quite
significant for a very massive particle as we shall see. Clearly the
single particle density of states diverges on the horizon if $s_{min} = 0$.
To obtain
the single string density of states we multiply $g(w,m)$ by $\rho(m)$
and integrate over the mass just as we did for the stress energy.

The entropy is given
in the canonical ensemble by forming $S = \beta U +\log Z$ where
$U = - {\partial \over \partial_{\beta} }\log Z$ and
$$
\log{Z} = -\int_{m_0}^{\infty} dm \int_{m s_{min}}^{\infty} dw \rho(m) g(w,m)
{1\over 2}\log\Biggl(
{
{1-e^{-\beta w}}
\over
{1+e^{-\beta w}}
}\Biggr)\eqno(3.3)$$
with $\beta$ set equal to $2\pi$. For massless states the entropy was
calculated by t'Hooft \NPrefmark{\thoo} and Susskind
and Uglum \NPrefmark{\suss} who showed that
$$S_{massless} = {A n_0\over 360 \pi s_{min}^2 }\eqno(3.4)$$
with $n_0$ related to the number of massless particles.
To calculate the entropy of massive states it is
convenient to define $E= w/s_{min}$ and $p = \sqrt{ E^2 - m^2 }$.
Then the single particle
density of states in Rindler space becomes
$$g(E,m) = {A\over (2\pi)^2}\Bigl( Ep + {m^2\over 2}
\log\bigl({ {E-p \over E +p}}\bigr)\Bigr)\eqno(3.5)$$
as compared with a Minkowski space density of
states that goes like $E p$. The partition function now becomes
$$
\log{Z} = -\int_{m_0}^{\infty}dm \int_m^{\infty} dE \rho(m)
s_{min}{A\over (2\pi)^2}\Bigl( Ep + {m^2\over 2}
\log\bigl({ {E-p \over E +p}}\bigr) \Bigr)
{1\over 2}\log\Biggl(
{
{1-e^{-\beta E s_{min}}}
\over
{1+e^{-\beta E s_{min}}}
}
\Biggr).\eqno(3.6)$$
For very massive string
states $E \approx m + {p^2 \over 2m}$ and a nonrelativistic approximation
is appropriate. Defining $v = p/E$ the density of states takes
the simplified form
$g(E,m) \approx {A\over (2\pi)^2} m^2
( {v\over {1- v^2}} - v- {1\over 3}v^3) \approx
{A\over (2\pi)^2}{2\over 3} m^2 v^3$ for massive nonrelativistic states. Note
that if the term involving the logarithm in (3.2) were not present
the density of states would only go like a single power of the velocity.

With the simplified density of states the partition function becomes
$$
\log{Z} = -\int_{m_0}^{\infty} dm \int_0^{\infty} dv \rho(m)
s_{min}{A\over (2\pi)^2}{2\over 3}m^3 v^4
{1\over 2}\log\Biggl(
{
{1-e^{
-\beta s_{min}m- \beta s_{min}{mv^2 \over 2}
}}\over{
1+e^{
-\beta s_{min}m- \beta s_{min}{mv^2 \over 2}
}}
}\Biggr).\eqno(3.7)$$
For very
massive particles the argument of the logarithm is near one, so
expanding the logarithm and integrating over the velocity $v$
we obtain
$$
\log{Z} = \int_{m_0}^{\infty} dm m_0^{n_s-1} m^{-n_s}e^{bm}
{1\over 4} s_{min}{A\over (2\pi)^2}m^3 \sqrt{\pi}
(\beta s_{min} m/2)^{-{5\over 2}}e^{-\beta s_{min} m}.\eqno(3.8)$$
The free energy $U = - {\partial \over \partial_{\beta}}\log Z$ is then given
by
$$
U = \int_{m_0}^{\infty}dm m_0^{n_s-1} m^{-n_s}e^{bm}
{1\over 4} s_{min}{A\over (2\pi)^2}m^3 \sqrt{\pi}
(\beta s_{min} m/2)^{-{5\over 2}}
(s_{min} m + {5\over 2\beta})e^{-\beta s_{min} m}.\eqno(3.9)$$
The entropy is then $S = \beta U + \log{Z}$ and performing
the integral over mass using the incomplete gamma function
$\Gamma(a,z) = \int_z^{\infty} dt t^{a-1} e^{-t}$ it can be expressed as
$$\eqalign{
S_{massive} =
{A\over s_{min}^2} {
\sqrt{\pi}\over 16\pi^2 (\beta /2)^{{5\over 2}}
}
\Biggl[
&\beta(m_0s_{min})^{{3\over 2}}
\bigl( m_0 (\beta s_{min}-b)\bigr)^{-{5\over 2}+n_s}
\Gamma\Bigl({5\over 2} - n_s, m_0 (\beta s_{min}-b)\Bigr)\cr
&+
{7\over 2} (m_0s_{min})^{{1\over 2}}
( m_0 (\beta s_{min}-b))^{-{3\over 2}+n_s}
\Gamma\Bigl({3\over 2} - n_s, m_0 (\beta s_{min}-b) \Bigr)
\Biggr]\cr}
.\eqno(3.10)$$
Now we take the inverse Hawking temperature $\beta = 2\pi$ and from
section two $s_{min}= {b\over 2}$. Noting
that $s_{min}$, $m_0^{-1}$ and $b$ are all of order $\sqrt{\alpha'}$ we
set and $m_0 = n_s'/s_{min}$ with $n_s'$ a number
of order one. The entropy due to the massive string states then becomes
$$S_{massive} = {A \over s_{min}^2} r(n_s,n_s')\eqno(3.11)$$
where
$$\eqalign{
r(n_s,n_s') = {1\over 16\pi^4}
\Biggl[
&2\pi(n_s')^{{3\over 2}}
( 2n_s'(\pi-1))^{-{5\over 2}+n_s}
\Gamma\Bigl({5\over 2} - n_s, 2n_s'(\pi-1)\Bigr) \cr
&+
{7\over 2} (n_s')^{{1\over 2}}
( 2n_s'(\pi-1))^{-{3\over 2}+n_s}
\Gamma\Bigl({3\over 2} - n_s, 2n_s'(\pi-1)\Bigr)
\Biggr]
\cr}
.\eqno(3.12)$$

The total entropy is the sum of that due to the
massless and massive states and is given by
$$S = S_{massless} + S_{massive} = {A\over s_{min}^2}
({n_0 \over 360\pi } + r(n_s,n_s') ),\eqno(3.13)$$
where $n_0$ is related to the number of massless states of the string theory
through $n_0= n_{b0} + {7\over 8}n_{f0}$ with $n_{b0}$ and $n_{f0}$
the
number of massless bosonic and fermionic modes.

Setting (3.13) equal to the Bekenstein-Hawking black hole entropy
$S_{BH}={A\over 4G}$
yields the condition
$${G\over 4} = s_{min}^2 /({n_0 \over 360\pi } + r(n_s,n_s') )\eqno(3.14)$$
or since $s_{min} = b/2$
$$G = b^2 /({n_0 \over 360\pi } + r(n_s,n_s') ).\eqno(3.15)$$
For all string theories $b$ is proportional to $\alpha'$ so that
(3.15) gives a relation between Newton's constant and the string tension.
Setting $b=n_s''\sqrt{\alpha'}$ we have:
$$G = \alpha'
{
{n_s''}^2 \over
{n_0 \over 360\pi } + r(n_s,n_s')
}.\eqno(3.16)$$
Therefore the relation between Newton's constant and the string tension is
fixed
by requiring the entropy of massless and massive string states to be equal
to the entropy of a black hole. For heterotic superstring theory compactified
to four dimensions $n_s=4$ and $b=(2+\sqrt{2})\pi\sqrt{\alpha'}$ so
$n_s''=(2+\sqrt{2})\pi$. For open superstring theory compactified to four
dimensions $n_s={5\over 2}$ and $b=2\sqrt{2}\pi\sqrt{\alpha'}$ so that
$n_s''= 2\sqrt{2}\pi$. In either case $m_0={n_s'\over s_{min}}={2n_s'\over b}$
can be chosen so that $n_s'$ is of order one. The massless contribution to
the entropy turns out to be much greater than that of the massive modes
whose major effect is to set the cutoff length $s_{min}$ as discussed
in section two. This being the case the relation between Newton's
constant and the string tension is of order $G \sim {10^5\over n_0} \alpha'$
where $n_0$ is related to the number of massless modes.

String theory itself relates Newton's constant to $\alpha'$ through
gauge and string coupling constants. These relations depend on the
type of string theory considered. For heterotic superstring theory
compactified to four dimensions \NPrefmark{\gros},
setting $\kappa_4 = \sqrt{8\pi G}$, the gauge
coupling to $g_4$, the string coupling to $g$, and the compactified six volume
to $V$, the relation is $2\kappa_4 = \sqrt{2\alpha'} g_4 =
{(2\alpha')^2\over \sqrt{V}} g$. In terms of Newton's constant we have
$G={g_4^2\over 16\pi}\alpha'= {g^2\over 2\pi V} {\alpha'}^4$. For open
superstrings \NPrefmark{\schw} the relations
are $\kappa_4\sim {\sqrt{V}\over \alpha'} g_4^2=
{8{\alpha'}^2\over \sqrt{V} }g^2$ which can be put in the form for Newton's
constant
$G \sim {Vg_4^4\over 8\pi{\alpha'}^3}\alpha' = {8g^4\over \pi V}{\alpha'}^4$.
The open superstring gauge coupling can be weakly coupled with weakly
coupled sigma model $V >> {\alpha'}^3$ and still be consistent with the
relation (3.16) derived by setting the entropy of the string states to the
Bekenstein-Hawking formula. All other couplings must either be strong or have
a strongly coupled sigma model $V << {\alpha'}^3$ to be consistent with
(3.16).

In \NPrefmark{\thoo} t'Hooft's studied the entropy of
a particle theory of fixed mass and infinite numbers of degrees of
freedom and found an unreasonably large value of the cutoff length.
Indeed requiring that the entropy agrees with the Bekenstein-Hawking
formula fixes
the cutoff length
$$s_{min} = {\sqrt{G} \over 2}\bigl({n_0 \over 360\pi }
+ r(n_s,n_s') \bigr)^{1\over 2}.\eqno(3.17)$$
Taking the limit $n_0 \rightarrow \infty$ tells us that
$s_{min} \sim \sqrt{n_0}$
if the entropy is to reproduce the Bekenstein-Hawking entropy.
Thus for an infinite
number of degrees of freedom the cutoff $s_{min}$ would have to move
infinitely far away from the horizon yeilding an unphysical picture
of a macroscopic black hole. String theory, on the other hand, has
an infinite number of degrees of freedom of increasing mass and a
fixed value for $s_{min} = b/2$ with a forbidden region only very
close to the horizon. This region corresponds to Planckian acceleration,
and it is physically reasonable for string effects to play a role there.

Our calculation of the black hole entropy in string theory
was done assuming the canonical ensemble.
There are concerns about the validity of using the canonical ensemble
because of the negative specific heat and loss of equilibrium
of both black holes and massive string
states. In this paper as in \NPrefmark{\suss} the area of the
horizon was taken to be $L^2$
with $L$ the limit of transverse coordinates in Rindler space. Rindler space
was identified with the spacetime of a very large black hole with horizon area
$(2GM)^2$ and, as in Rindler space, was taken to be static. For a very large
black hole, the variation in the horizon area with time is so slight that the
identification with Rindler space is appropriate and the deviation from
equilibrium relatively small. However for small evaporating black holes
the horizon area is rapidly varying,
with a metric deviating strongly from Rindler space.
In this regime a microcanonical counting of states is
more appropriate procedure far from equilibrium.

A microcanonical description
leads to a number of states $\sigma(W) = \exp\bigl(S(W)\bigr)$ with $W$
related to the mass of the black hole and
$$
\sigma(W) = \sum_N {1\over N!} \prod_{i=1}^N \int_{m_0}^{\infty} dm_i
\int_{m_is_{min}}^{\infty}dw_i
\rho(m_i)g(w_i,m_i) \delta(W - \Sigma_i w_i).\eqno(3.18)$$
Here $N$ represents the number of strings and ${1 \over N!}$ ensures the
correct statistics. The single particle density of states $g(w,m)$ can
be that relevant to a small evaporating black hole out of equilibrium.
A similar microcanonical description has been used to describe
string theories at high energy density \NPrefmark{\frau} \NPrefmark{\turo}
where a single massive string state
can carry almost
all the energy and represent a nonequilibrium configuration. In this way
massive string states can also lead to a negative specific heat
\NPrefmark{\bowick}. However,
in the case of a very large black hole studied in this paper
the canonical ensemble is valid because the limiting Hawking
temperature is still less by a factor of ${1 \over \pi}$ from the Hagedorn
temperature above which thermodynamic quantities can diverge.

\noindent{\bf IV. Conclusion}

In this paper we have examined the arguments leading to a maximum acceleration
in string theory and an infinite wall of stress a finite distance outside the
horizon. We multiplied the single massive particle density of states of
\NPrefmark{\thoo,\suss}
by the number of string modes at a given mass and summed
over all masses to obtain a single string density of states
in Rindler space or about a very large black hole. We found that a
cutoff on the density
of states was introduced by the infinite wall of vacuum stress and maximum
acceleration. The entropy of the string excitations was computed
using the canonical ensemble at the Hawking temperature and was finite
because of the natural cutoff. The massive string
contribution preserved the relation
that the entropy is proportional to the area of the horizon divided by the
cutoff squared. We found that the string entropy agreed with the
Bekenstein-Hawking
formula if a particular relationship existed between Newton's constant
and the string tension, and then compared this with similar relationships
found in various string theories. Finally we discussed the validity of
the canonical ensemble for evaporating black holes and massive string states.

The basic point is that there is a region
outside the horizon of a black within which a string cannot remain
stationary. Likewise, by the equivalence principle, there is a region
outside the horizon of Rindler space within which a string cannot
accelerate into. This suggests that an effective cutoff is
introduced on the boundary of a region slightly away from the horizon.
This can cut off the divergence in the single particle density
of states and yield a finite answer for the entropy. t'Hooft's result
that the brick wall cutoff should move infinitely away from the horizon
in a theory with infinite numbers of degrees of freedom is avoided in
string theory due to the infinite set of states of arbitrarily high mass.
Further studies using a fundamental description of the string propagator
\NPrefmark{\cohe}, instead of the sum over modes approach
we have taken here, are necessary to place the
entropy calculation on a sound footing. In general,
string calculations have features like modular invariance which are not
obvious in a sum over field theories and are important for a geometric
understanding of the partition function and entropy.

\noindent{\bf Acknowledgments}

I wish to thank Z.Qiu and C.Thorn for useful discussions as well as
L.Susskind and G.t'Hooft for an inspiring set of lectures.
\refout
\endpage
\noindent{\bf Figure Captions}

\noindent Figure 1. Representation of the structure of Rindler space in
string theory. The dashed line indicates the $s= b/2= s_{min}$ boundary
inside
of which the vacuum energy diverges and accelerations exceed
$a= 2/b = a_{max}$. The outside curved line is a trajectory of
an accelerated observer with $a < a_{max}$ and the
dark 45 degree lines show the horizon $x = |t|$.

\bye